\documentclass[%
 reprint,
superscriptaddress,
 amsmath,amssymb,
pra,
]{revtex4-2}
\usepackage{url}
\usepackage{braket}
\usepackage{graphicx}
\usepackage{dcolumn}
\usepackage{bm}
\usepackage{color,soul}
\usepackage{hyperref}
\hypersetup{
    colorlinks=true,
    linkcolor=blue,
    filecolor=magenta,      
    urlcolor=cyan,
}

\newlength{\onecolfig}
\newlength{\twocolfig}
\usepackage{physics}

\newcommand{\ps}{\,\mathrm{ps}}

\newcommand{\us}{\ensuremath{\,\mathrm{\mu s}}}

\newcommand{\s}{\,\mathrm{s}}

\newcommand{\yb}{\ensuremath{^{171}\mathrm{Yb}^+\,}}

\newcommand{\uuKet}{\ensuremath{\ket{\uparrow \uparrow}}\,}

\newcommand{\udKet}{\ensuremath{\ket{\uparrow \downarrow}}\,}

\newcommand{\up}{\ensuremath{\ket{\uparrow}}\,}
\newcommand{\dn}{\ensuremath{\ket{\downarrow}}\,}
\newcommand{\pls}{\ensuremath{\ket{+}}\,}
\newcommand{\mn}{\ensuremath{\ket{-}}\,}

\newcommand{\psz}{\ensuremath{\sigma_z}}

\begin{document}

\preprint{APS/123-QED}

\title{Quantum computing hardware in the cloud:\\ Should a computational chemist care?}

\author{Alessandro Rossi}
\email{alessandro.rossi@strath.ac.uk}
\affiliation{
Department of Physics, SUPA, University of Strathclyde, Glasgow G4 0NG, United Kingdom}
\affiliation{National Physical Laboratory, Hampton Road, Teddington TW11 0LW, United Kingdom
}
\author{Paul G. Baity}%
\affiliation{
James Watt School of Engineering, University of Glasgow, Glasgow G12 8LT, United Kingdom
}
 \author{Vera M. Sch\"{a}fer}
 \affiliation{
Department of Physics, University of Oxford, Clarendon Laboratory, Parks Road, Oxford OX1 3PU, United Kingdom
}
 \author{Martin Weides}%
\affiliation{
James Watt School of Engineering, University of Glasgow, Glasgow G12 8LT, United Kingdom
}%


\date{\today}

\begin{abstract}
Within the last decade much progress has been made in the experimental realisation of quantum computing hardware based on a variety of physical systems.  Rapid progress has been fuelled by the conviction that sufficiently powerful quantum machines will herald enormous computational advantages in many fields, including chemical research. A quantum computer capable of simulating the electronic structures of complex molecules would be a game changer for the design of new drugs and materials. Given the potential implications of this technology, there is a need within the chemistry community to keep abreast with the latest developments as well as becoming involved in experimentation with quantum prototypes. To facilitate this, here we review the types of quantum computing hardware that have been made available to the public through cloud services. We focus on three architectures, namely superconductors, trapped ions and semiconductors. For each one we summarise the basic physical operations, requirements and performance. We discuss to what extent each system has been used for molecular chemistry problems and highlight the most pressing hardware issues to be solved for a chemistry-relevant quantum advantage to eventually emerge.  
\end{abstract}

\maketitle


\section{\label{sec:intro}Introduction}

This year marks exactly 40 years since Richard Feynman famously said~\cite{Feynman}: “Nature isn't classical, dammit, and if you want to make a simulation of nature, you'd better make it quantum mechanical, and by golly it's a wonderful problem, because it doesn't look so easy”. On the one hand, the visionary physicist anticipated the possibility (and the inherent difficulty) of building a new type of computing apparatus operating according to the laws of quantum mechanics. On the other hand, he had immediately identified one of its most useful areas of application, i.e. simulations of chemical and physical systems.\\\indent 
Computational chemists will indeed benefit from future quantum computers for calculations of molecular energies to within chemical accuracy, defined to be the target accuracy necessary to estimate chemical reaction rates at room temperature ( $\approx  1$~kcal/mol)~\cite{elfving2020quantum}. Fully-fledged, error-free quantum systems will enable predictions and simulations that are not possible today in terms of both accuracy and speed. This could have a revolutionary impact on the design of drugs, catalysts and materials by allowing computational methods to replace lengthy and expensive experimental procedures. Unfortunately, we are still in the infancy of the development of quantum computing technology and a machine that provides a quantum advantage in molecular chemistry over classical super-computers has not emerged yet. However, the progress in handling increasingly complex molecular and material chemistry has been relentless. Small-scale quantum machines developed by academic or corporate research centres have been initially used to simulate simple diatomic or triatomic molecules made up of just H and He atoms~\cite{Colless,Shen,OMalley2016}. Recently, more powerful quantum computers have been used to simulate larger compounds containing N, Li and Be atoms~\cite{Arute2020_1,Kandala2017,Kandala2019}. Although these studies do not show a clear advantage in using quantum computing over the conventional computational methods that have been used for their validation, they do indicate that hurdles are being tackled and viable ways forward are becoming available.\\\indent  
The major impediments that currently stifle quantum computers are limits to the number of computational units and computational errors. The units of quantum information are called quantum bits (qubits) in analogy with the binary bits of classical computers. Quantum algorithms for chemical calculations use qubit-based Hamiltonians to map molecular many-body Hamiltonians and evaluate the system wavefunction through repeated sampling of the qubit register states
~\cite{Lanyon2010,Aspuru-Guzik1704,RevModPhys.92.015003}. One particular algorithm, namely the variational quantum eigensolver (VQE)~\cite{peruzzo2014}, has acquired prominence because it alleviates the computational burden on today's limited quantum machines by using a classical co-processor to support the calculation. To date, the most advanced VQE simulations have mapped just 24 molecular orbitals onto 12 qubits~\cite{Arute2020_1}, a relatively easy feat for traditional computers. In order to calculate the energy ground state of more complex systems with chemical accuracy, it is expected that the number of qubits available will need to increase by orders of magnitude. A recent estimate~\cite{elfving2020quantum} indicates that more than 1500 orbitals are required for a VQE calculation that could outperform classical super computers.\\\indent  
The other hurdle to consider is that qubits are error-prone due to noise-limited phase coherence, with inherent challenges in reading and writing their states properly (qubit fidelity). Ultimately, there is a limit to the number and duration of operations (qubit gates) that a quantum computer can carry out before error propagation leads to computational failure. Quantum error correction (QEC) schemes to correct these errors have been identified~\cite{Fowler2012,Ofek2016}. 
The main drawback is that QEC leads to hardware aggravation, given that several physical qubits are required to realise a single error-corrected ``logical'' qubit. Some estimates based on realistic qubit noise levels conclude that the ratio of physical to logical qubits to reach fault tolerant machines could be at least $1$,$000:1$~\cite{martinis2015}.\\\indent 
It is, therefore, evident that, to approach quantum chemistry simulations in a meaningful way, quantum computers with millions of physical qubits will be required, if one has to accurately map thousands of spin-orbitals. By contrast, today's quantum computers rely on a small number of noisy qubits (less than 100 at present) because the ability to manufacture, interconnect and error-correct qubits on larger scales is not yet sufficiently developed. This is why quantum machines are presently dubbed NISQ (Noisy Intermediate-Scale Quantum)~\cite{Preskill2018quantumcomputingin}. An important figure of merit for NISQ systems is called quantum volume (QV)~\cite{Cross2019}, which combines in one convenient metric the number of qubits available, how extensively they are interconnected, and their gate fidelity. A larger QV indicates that more complex quantum algorithms can be successfully run. This metric clearly shows that, to increase the computational power, is not sufficient to build machines with more qubits if these remain affected by high levels of noise. Hence, the challenge of improving quantum computing power is a coordinated effort in scaling up qubits, making them as interconnected as possible, and reducing the error rates.\\\indent
NISQ computers come in a variety of hardware implementations. Different from classical computers for which the Central Processing Unit (CPU) is invariably made with silicon integrated technology, Quantum Processing Units (QPU) can also be realised with superconductor microchips, ions or neutral atoms trapped in a vacuum, and on-chip photonic waveguides. Different technologies present different trade-offs in terms of number of qubits, phase coherence time, qubit fidelity, connectivity etc.  Here, we are going to focus on a specific subset of quantum hardware types. Specifically, we will look at digital programmable QPUs, as opposed to adiabatic or analog systems~\cite{albash2018,Hauke_2020}. Among these, we shall discuss computers available to the general public through cloud services. On the one hand, being available to the public, and not just to specialised quantum developers, indicates these systems have reached superior maturity. On the other hand, we feel that a description of how these systems operate at hardware level will benefit the reader who may have to navigate through offers and subscription packages to identify the most relevant service for the computational chemistry application of interest. This may indeed become a daunting task without prior knowledge given the pace with which these services are becoming available and compete to acquire large customer bases. Global corporations offering cloud access include Google, IBM, Microsoft and Amazon. We shall limit our discussion to three types of hardware in the cloud: superconductor-, ion trap- and silicon-based quantum computers. For each one of these systems we discuss how qubits are physically embodied, initialised, read and manipulated. We will describe the operational requirements and the main performance parameters of each implementation. We will provide some use cases relevant to quantum chemical simulations to exemplify the usefulness of different machines in relevant contexts. The remainder of this Article is organised as follows. Superconductor devices are described in Section~\ref{sec:super}, ion trap systems in Section~\ref{sec:ion}, and a silicon processor in Section~\ref{sec:silicon}. These technologies are compared in Section~\ref{sec:disc}, and finally an outlook for future developments is discussed in Section~\ref{sec:conc}.

\section{\label{sec:super}Superconducting quantum computers}

\begin{figure}[]
\includegraphics[scale=0.5]{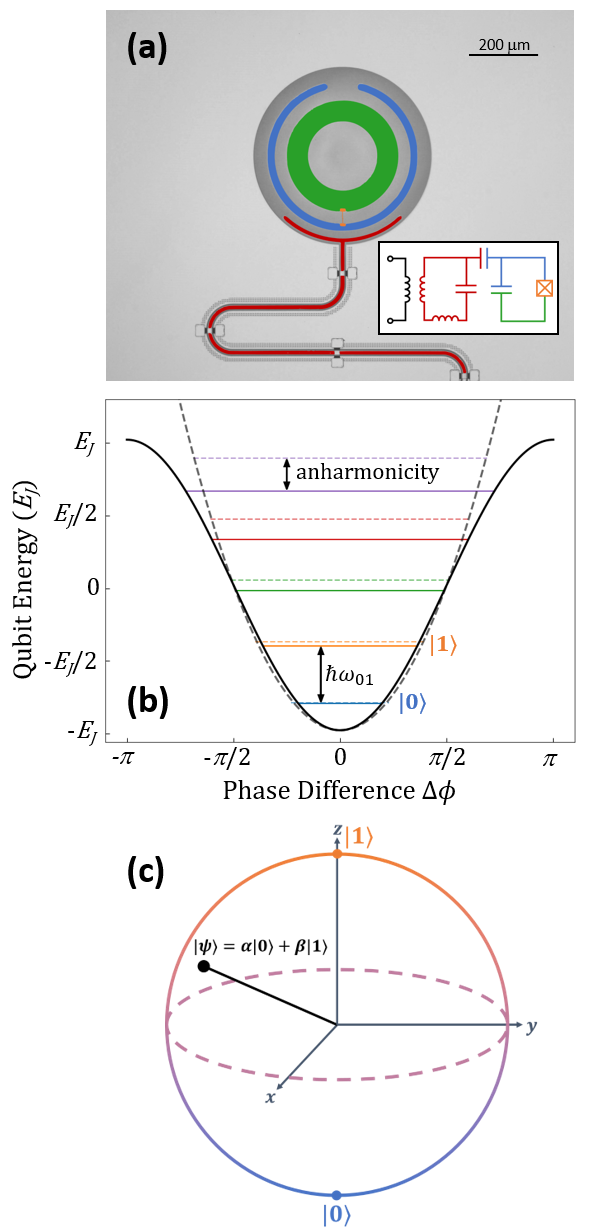}
\caption{(a) Concentric transmon qubit design from Ref.~\cite{Stehli2020} and (inset) its equivalent circuit diagram. Two superconducting islands (green and blue) are shunted by a small Josephson junction bridge (orange). The qubit state is read out using a coplanar waveguide resonator (red). This readout resonator is inductively coupled to a signal line (black). (b) The states of the transmon qubit are determined by the sinusoidal potential (black solid line) of the Josephson junction. Solved in the phase basis ($\Delta\phi$), the Eigen energies (solid colored lines) can be approximated by a harmonic oscillator (dashed lines, respective colors) whose degeneracies are lifted by first order corrections from capacitive charging energy on the junction \cite{Koch2007, Wilkinson2018, Krantz2019}. (c) Diagram of the Bloch sphere. The ground $\ket{0}$ and first excited $\ket{1}$ states are used to define the qubit's logical state $\ket{\psi}$, which is a linear combination of $\ket{0}$ and $\ket{1}$ with respective complex amplitudes $\alpha$ and $\beta$. $\ket{\psi}$ can be manipulated by voltage pulses and gating operations and read out by projection onto a specified measurement basis.}
\label{fig:transmon}
\end{figure}

Superconducting (SC) circuits are the most widely used systems for quantum computing. Many industry leaders, such as Google, IBM, and Rigetti, use superconducting quantum circuits to realize their quantum computers. Qubits implemented on superconducting devices fulfil the requirements \cite{Divincenzo2000} for scalable quantum computing, and therefore micron-sized quantum circuits and associated integrated-circuit processing techniques can be scaled up when implemented using superconducting quantum technologies. Whereas trapped ion and silicon devices control and read (sub-)atom scale components as their quantum systems, in SC circuits information is encoded into a \emph{macroscopic} quantum state of the condensate of paired electrons (so-called Cooper pairs), which collectively participate in a charged superfluid state with a wave function $\Psi(\vec{r},t) = |\Psi(\vec{r},t)|e^{i\phi(\vec{r},t)}$ \cite{BCS, Tinkham}. Here, the wave function parameters $|\Psi(\vec{r},t)|^2$ and $\phi(\vec{r},t)$ describe the density of Cooper pairs and complex phase of the condensate as a function of position $\vec{r}$ and time $t$.
Superconducting qubits, such as the one shown in Fig.~\ref{fig:transmon}(a), consist of islands of superconducting material, such as aluminium, connected by one or more Josephson junctions \cite{Tinkham}, which are nm-thin insulating barriers made from e.g. aluminium oxide. 
The current $I$ passing through the Josephson junction depends on the phase difference $\Delta\phi$ between the superconductors at either side of the junction by the relation $I = I_0 \sin(\Delta\phi)$, where $I_0$ is the largest supercurrent supported by the junction. When a voltage difference $V$ occurs across the junction, $\Delta\phi$ changes as $\frac{d\Delta\phi}{dt}=2eV/\hbar$ \cite{Josephson, Tinkham}, where $e$ is the electron charge and $\hbar$ is the reduced Planck's constant. This time dependence leads to non-linear resonance behavior with quantized states that are determined by flux, charge, and phase degrees of freedom \cite{Clarke2008}. 

The effective circuit diagram of a superconducting qubit is shown in the inset of Fig.~\ref{fig:transmon}(a) and can be described by the Hamiltonian \cite{Koch2007,Krantz2019}
\begin{equation}
H=4E_C (\Delta n)^2-E_J \cos(\Delta\phi), \label{transmonhamiltonian}
\end{equation} 
where $E_J$ is the energy of the current passing through the junction and $E_C$ is the capacitive charging energy between the two superconducting islands. 
Quantum states are usually determined in either the basis of $\Delta \phi$, as shown in Fig.~\ref{fig:transmon}(b), or capacitive charge number $\Delta n$, depending on the relative strengths of $E_J$ and $E_C$. Similar to the conjugate variables of position and momentum, $\Delta \phi$ and $\Delta n$ have a non-zero commutation $[\Delta \phi, \Delta n] = i$ \cite{Krantz2019} and uncertainty relation $\sigma_{\Delta\phi}\sigma_{\Delta n} \gtrsim 1$ \cite{Tinkham}. In either basis, quantum states are approximated by a harmonic oscillator in which degeneracy is lifted by the non-dominant energy term. This degeneracy allows for the differentiation of qubit states. Superconducting qubits are typically operated at the transition between the ground $\ket{0}$ and first excited $\ket{1}$ states.


One of the benefits of superconducting qubits is the ability to engineer a wide range of operational parameters by tuning the parameters $E_J$ and $E_C$ through intentional design choices. Perhaps the most widely used design choice is to have $E_J/E_C \sim 10^2$. This is the so-called transmon qubit design \cite{Koch2007}, which has been widely used by both academic and industry leaders to realize quantum computers. This ratio of $E_J/E_C$ creates an exponential cutoff for charge fluctuations, leading to longer lifetimes. Since $E_J$ is large compared to $E_C$, the quantum Eigen states are determined in the $\Delta \phi$ basis as shown in Fig.~\ref{fig:transmon}(c). The Eigen energies have an approximate $\sqrt{8E_J E_C}$ separation, while first order corrections on the scale of $E_C$ create the essential anharmonicity between energy levels that is required for two-state control \cite{Wilkinson2018}. Therefore the transmon's $E_J/E_C$ ratio is large to reduce charge noise but small enough to prevent excitation beyond the first excited state. 

\subsection{Qubit initialization \& readout}


The qubit is a quantum mechanical two-level system with logical states $\ket{0}$ and $\ket{1}$, in analogy to a classical bit. Without any external or thermal excitation, the superconducting qubit state $\ket{\psi}$ relaxes into the $\ket{0}$ state. Under a resonant drive $\ket{\psi}$ will oscillate between $\ket{0}$ and $\ket{1}$ as a superposition on the surface of the Bloch sphere, shown in Fig.~\ref{fig:transmon}(c). The measured period of these so-called Rabi oscillations is used to calibrate the applied microwave drives for qubit control. Reading the state of the qubit requires projecting $\ket{\psi}$ onto the quantization axis. Information about the probability distribution along other directions is obtained by fast rotations of the axis in question onto the quantization axis and subsequent measurement. Fast, in this context, means that the pulse length is short compared to the respective decoherence times. By measuring in quick succession, the qubit state can be inferred from the probabilities of the measurement results.

Superconducting qubit states are usually determined using dispersive readout \cite{Krantz2019}, where $\ket{\psi}$ is not measured directly but is inferred from measurements of a coupled photon resonator. The interaction between the resonator and qubit shifts the effective frequency of the resonator by an amount dependent on the projection of $\ket{\psi}$. Therefore, the qubit state can be inferred by measurements of the resonator frequency. However the resonator frequency, $\omega_{r}$, must be detuned from the qubit frequency, $\omega_{01}$, to prevent measurements from interfering with the qubit state. The detuning frequency, $\Delta = \omega_{01} - \omega_{r}$, is greater than the coupling rate, $g$, between the qubit and resonator to ensure that energy is not coherently exchanged between the qubit and resonator. This condition prevents a measurement from affecting subsequent measurements (quantum non-demolition). This control scheme does have a drawback: since the qubit is coupled to the resonator, noise within the resonator can cause arbitrary phase decoherence in the qubit. Therefore, measurement signals used to probe the resonator frequency must be attenuated and filtered to reduce noise and ensure qubit fidelity.

\subsection{Qubit manipulation}

The microwave tones used for qubit manipulation are referred to as \emph{gates} or \emph{pulses}. Qubit manipulation is achieved with a heterodyning technique, where the pulse signal is generated by a mixer, modulating a baseband signal of a local oscillator operating close to the desired frequency, with an envelope at lower frequency. The envelope is generated by fast digital-to-analog converters, which generate both components of the manipulation or readout pulses in the respective baseband. For readout the returning microwave signal from the readout resonator gets down-converted with the same local oscillator used for the up-conversion, yielding the demodulated baseband signal. After low pass filtering to suppress leakage of the carrier frequency and further amplification, the signal is digitized by an analog-to-digital converter card. Fourier transformation of the incoming signal for both quadratures gives the complex scattering parameter and in the case of dispersive readout, the state of the qubit.

Single qubit gates correspond to rotations of a Bloch vector about some axis of the Bloch sphere while multi-qubit gates take two or more qubits as input to manipulate at least one qubit state. Multi-qubit gates require entangling two or more qubits together, while a series of pulses is applied to one or multiple qubits \cite{Kjaergaard2020}. Since the qubits are entangled, the readout of one qubit can be manipulated via a second, ancillary qubit. An example is the Controlled NOT gate (or CNOT gate), where a target qubit state is flipped if and only if a second, control qubit state is $\ket{1}$. Quantum logic gates are the fundamental basic quantum circuit building block, operating on a small number of qubits (usually one or two). However, it should be noted that due to the planar structure of superconducting circuitry, connectivity between qubits is currently limited to nearest-neighbor interactions. This imposes constraints on gating capabilities, as operations between non-neighboring qubits cannot be performed. There are current aims to realize 3D-integrated superconducting circuits \cite{Rosenberg2017,Yost2020}, which will allow additional connections for beyond-nearest-neighbor interactions, overcoming the current limitations. 

\subsection{Operational conditions \& performance indicators}
Like other quantum systems, calculations are limited by the longitudinal and transverse relaxation times, $T_1$ and $T_2$. With current technology, decoherence rates below 1 MHz can be achieved \cite{Krantz2019}, allowing for the creation and manipulation of single or multiple quantum excitations in superconducting qubits with fast (nanosecond) control. Improvements to qubit lifetimes have been achieved primarily through qubit design, improvements in fabrication quality, and material selection. For current systems, lifetimes are long enough to ensure computational fidelity. Indeed, for the very best SC QPUs in the cloud, 1- and 2-qubit gate fidelities exceed 99\% with qubit readout errors in the range of 2-3\%.

Regardless of design, qubits must be operated well under the superconducting transition temperature $T_c$. Furthermore, since SC qubits are strongly coupled to their environment and readout circuitry, thermal and electromagnetic noise should be reduced as much as possible. Therefore, qubits are usually measured and operated at $T=10$~mK in dilution refrigerators with magnetically shielded environments. As mentioned previously, measurement lines are also typically thermalized and attenuated to reduce noise. The need for cryogenic environments currently imposes a limitation on the size of SC quantum computers, since each measurement line leaks heat into to the system and decreases the effective temperature of the refrigerator. This limitation can be overcome by implementation of cryogenic processing and multiplexing of classical signals. Two potential platform solutions are the cryogenic complementary metal-oxide-semiconducting (Cryo-CMOS) \cite{Homulle2016,Charbon2019,Pauka2021} and rapid single flux quantum (RSFQ) \cite{Likharev1991} hardware platforms, which can serve as low-temperature interfaces between classical and quantum systems.

Another limit is the speed at which qubits can be operated. At high frequencies, superconductivity breaks down as single electrons are excited out of the superfluid \cite{Tinkham}. The presence of these quasiparticles leads to dissipation and decoherence, and thus qubits are typically designed to operate at frequencies $\omega_{01}\ll k_B T_c/\hbar$. For aluminium with $T_c = 1$~K, qubits are typically designed to operate at $\omega_{01} < 20$~GHz. Additionally, while the macroscopic nature of superconducting qubits allows for customization of qubit parameters, this benefit comes with a drawback in producing identical qubits, as small deviations in fabrication uniformity can be difficult to control.

\subsection{Use case}
\label{sec:super:use}
Superconducting quantum circuits have been used to simulate many physical systems. 
Spin systems have been a particular focus for quantum simulation through both analog \cite{Guo2019,Xu2020,Fitzpatrick2017,Harris2018} and digital \cite{Salathe2015,Smith2019} methods. However, with regard to digital simulations, a recent study \cite{Smith2019} performed on an IBM QPU has concluded that the current state of SC quantum computers is too error-limited to produce dependable quantitative results for larger (six spins or more) systems.

Chemical binding energies of molecules have been calculated using VQEs
\cite{OMalley2016, Kandala2017, McCaskey2019, Karalekas2020} implemented on SC circuits. The VQE method has had relatively good success in determining binding energies of H$_2$, LiH, BeH$_2$, NaH, KH, and RbH. Due to the circuitry scale, these studies consider only a limited number of basis states (e.g. spin orbitals), allowing for a comparison to the exact, diagonalized solutions. In this context, the calculated results are in good agreement with theoretical expectations. More recently, binding energies of hydrogen chains up to H$_{12}$ have been modeled using Google's Sycamore QPU \cite{Arute2020_1}. However, it should be noted that several postprocessing techniques were required to mitigate errors in the raw results and achieve quantitative chemical accuracy for bonding energies. This work also simulated diazene (H$_2$N$_2$) isomerization energies for converting cis-diazene to trans-diazene, marking the first time a chemical transition has been modeled on a quantum computer. Therefore, despite the limitations from noise and basis size, digital simulations on SC QPUs show promise for chemical simulations.


\section{\label{sec:ion}Quantum computing with trapped ions}
Trapped ions \cite{Wineland1998,Haffner2008} were one of the first platforms proposed for building a quantum computer as they form a natural representation of an ideal qubit: all ions are identical by nature, their high degree of isolation from the environment leads to excellent coherence times and interaction with radio-frequency (rf) and laser light allows for high-fidelity gate operations.
Qubits are encoded in the electronic states of individual ions trapped by electric fields in an rf Paul trap. 
Two-dimensional traps can be micro-fabricated on silicon chips, called surface traps, and can contain multiple trapping and interaction zones as well as integrated microwave and laser access \cite{Chiaverini2005,Revelle2020,Mehta2020,Niffenegger2020}.
Interaction of the electronic states of neighbouring ions is negligibly small \cite{Kotler2014}, but ions are strongly coupled via their motion which can be exploited to create entanglement between different ions necessary for multi-qubit gates \cite{Monroe1995}.
Ions are confined in long chains, within which all ions can interact with each other. 
Chains can be split and merged, and ions can be moved across the chip between different zones, providing large flexibility of connections \cite{Blakestad2011,Bowler2012,Walther2012}.
Many different elements are used as ion species, but all ions are typically singly-charged and have a single remaining valence electron. 
Popular choices of ion are $\mathrm{Yb}^+$, $\mathrm{Ca}^+$ and $\mathrm{Be}^+$ \cite{Debnath2016b,Baldwin2020,Ballance2016,Erhard2019,Gaebler2016}. 
Qubit states can either both be encoded in ground-state levels (hyperfine- and Zeeman-qubits \cite{Harty2014,Ballance2015}) with transition frequencies in the rf range, or with the excited state encoded in a meta-stable state (eg. $\mathrm{D}_{5/2}$) leading to optical transition frequencies \cite{Benhelm2008}.
Different properties of the atomic species affect the qubit performance. 
For example, some hyperfine qubits are robust to magnetic field noise, which is the main source of decoherence in trapped ion qubits, and therefore have greatly enhanced coherence times \cite{Harty2014,Wang2020}.
Other important factors are the existence of low-lying D manifolds, which can assist readout but cause errors due to scattering in laser gate operations; the ions' mass where lighter ions allow faster gates; excited state lifetimes for optical qubits; and transition frequencies depending on the availability of suitable lasers.\\\indent 
Scaling up devices from tens to thousands or millions of qubits is arguably the biggest challenge in realising a quantum computer.
The trapped ion community pursues several paths towards scalability.
In the quantum charge-coupled-device (QCCD) architecture \cite{Kielpinski2002a,Wineland1998} ion chains are broken up into smaller groups in individual zones, instead of forming a single long string.
For scalability beyond a single chip proposals include connecting separate traps via photonic links \cite{Monroe2013a,Nigmatullin2016a,Stephenson2020} and shuttling of ions across arrays of chips \cite{Lekitsch2017}.
Another important ingredient for scalability is the simultaneous use of different ion species, which allows sympathetic cooling of ions without affecting the electronic state of the logic and memory qubits \cite{Kielpinski2000} and better spectral isolation for ion-photon entanglement.
Strings of ions can be split, merged and shuttled between different zones with negligible effect on the spin state and coherence, but a slight increase in ion temperature \cite{Blakestad2011,Bowler2012,Walther2012}.
While ion traps can be operated at room temperature, their performance is enhanced at cryogenic temperatures due to a reduction in heating rate and an increase in ion lifetime.
Cooling down to $\sim10\mathrm{K}$ with liquid helium cryostats suffices for this purpose.\\\indent
Trapped ions have the longest coherence times of all contending platforms for building a quantum computer. Even though their individual operations are slower than in solid state systems, they still possess a superior ratio of gate operation time to coherence time, which ultimately results in record single- and two-qubit gate fidelity.
While technology and infrastructure for solid-state systems is more mature than laser technology due to developments made for classical computer chips, rapid progress in the stability, miniaturisation, and integration of laser and ion trap systems has been achieved in the last few years due to the influx of resources and increase in demand. 
Trapped ion quantum computers also benefit from the absence of noisy direct environments which are present in solid state systems, and the high degree of connectivity and flexibility of connections in trapped ion systems.
Remaining challenges are to reduce gate errors for larger numbers of qubits, which tend to increase with the number of ions, and to improve automatisation, robustness and crosstalk for building larger devices.
Further research is also required in trap fabrication, as one of the major gate error sources stems from anomalous heating of the ion crystals, thought to be caused by surface effects on the ion trap electrodes \cite{Hite2017,Sedlacek2017,Boldin2018,Noel2019}.

\subsection{Qubit readout, initialisation and cooling}

Qubits are read out via state-dependent fluorescence detection. 
All ion species used for quantum computing have a short-lived excited state that predominantly decays back into the qubit ground state manifold. 
For optical qubits and some hyperfine qubits the qubit frequency is sufficiently large that the fluorescence laser only couples to one of the qubit states, the `bright' state. 
Together with selection rules preventing decay from the excited state into the opposite `dark' qubit state, this allows direct fluorescence readout.
For qubits without direct state selectivity of the fluorescence laser, the dark state is transferred into a `shelf' state that does not couple to the fluorescence laser and the excited state.
Ion-position resolved fluorescence can be detected with arrays of photomultiplier tubes or avalanche photodiodes, on an electron-multiplying charge coupled device camera \cite{Myerson2008}, or with superconducting nanowire single-photon detectors integrated into the trap chip \cite{Todaro2020}.
Fluorescence can be collected over a fixed time-bin and analysed with threshold or maximum likelihood algorithms, or with real-time analysis and adaptive readout duration. 
With sufficiently low background counts and high photon collection and detection efficiency, real-time analysis achieves the same fidelities as fixed-time threshold analysis, but is considerably faster \cite{Noek2013b,Todaro2020}.

Qubit initialisation is performed via optical pumping, using the same excited states as for fluorescence readout.
Either frequency or polarisation selectivity is used to ensure that population is excited out of all ground states apart from the target initial state.
Different states can be prepared by applying a sequence of single qubit operations after optical pumping.
 
For optimum gate fidelities ion crystals need to be cooled close to their motional ground states, which is performed with laser cooling.
Typically ions are continuously Doppler cooled during idle time.
Before an experiment resolved-sideband cooling (RSBC) is used to further cool relevant motional modes to an average motional mode occupation of $\bar{\mathrm{n}}\lesssim 0.1$.
Alternatively electromagnetically-induced transparency cooling can be used to cool all modes simultaneously \cite{Lechner2016}.
While considerably faster than RSBC, especially for larger ion strings, the final temperature reached is slightly higher.


\subsection{Qubit manipulation}

Single qubit gates can be driven directly using rf in Zeeman- and hyperfine-qubits, or using a narrow-linewidth laser to drive the quadrupole transition in optical qubits.
Alternatively a pair of lasers which are far detuned from the excited state and have the qubit frequency as their frequency difference can be used to drive qubit rotations via two-photon Raman transitions.
Rotations around the z-axis can be performed trivially by propagating the phase of all future operations.
The phase is defined by a direct digital synthesis frequency source that is either applied directly on the ions as rf or controls the frequency, amplitude and phase of the laser beams via an acousto-optic modulator (AOM).
Rf operations couple only very weakly to the motion due to their low photon energy and can already be performed at Doppler-cooled temperatures at very high fidelities \cite{Harty2014}.
They also have superior phase stability compared to lasers and can easily be integrated into surface traps, but are harder to address onto single ions.

Multi-qubit gates create entanglement between different qubits and require the ions' motion as a bus of interaction between the ions.
There are different schemes for entangling gates, with the most established ones being the closely related M\o lmer-S\o rensen (MS) gate \cite{Sorensen1999} and the \psz\ geometric phase (ZGP) gate\cite{Leibfried2003}. 
Both create a spin-dependent force on the ions; the MS gate in the \pls,\mn basis and the ZGP gate in the \up, \dn basis.
This force leads to motional excitation and displacement for one spin parity combination (eg \udKet) but not the other (eg \uuKet).
Displaced spin states acquire a phase which ultimately leads to the entanglement.
The propagator of a two-qubit gate with these schemes is diag(1,i,i,1), which corresponds to a controlled-PHASE gate. 
This gate can be transformed into a CNOT gate via additional single-qubit operations.
Both gate mechanisms are first-order insensitive to the ion temperature, which makes them more robust and is an important factor in the high fidelities achieved.
ZGP gates cannot be performed directly on the low-decoherence clock qubits, but are insensitive to the absolute magnetic field offset.
Two-qubit gates have been performed both with lasers \cite{Benhelm2008,Ballance2016,Gaebler2016,Baldwin2020b} and rf \cite{Ospelkaus2011,Harty2016,Weidt2016,Zarantonello2019,Srinivas2021} as well as between ions of different elements \cite{Ballance2015,Tan2015,Hughes2020,Bruzewicz2019}.
Due to the weak motional coupling rf multi-qubit gates are considerably slower than laser gates.
Gates can be performed globally on all ions in a string simultaneously or addressed locally to a specific subset of ions \cite{Erhard2019,Debnath2016b}.

\subsection{Performance indicators}

Coherence times in trapped ions are $T_2^*=50\s$ on magnetic-field insensitive clock qubits \cite{Harty2014} and reach over an hour by employing dynamical decoupling sequences \cite{Wang2020}. 
State-preparation and measurement errors are $\varepsilon <1\cdot 10^{-3}$, with a mean duration of  $46\us$ \cite{Todaro2020,Noek2013b,Myerson2008}, where $\varepsilon\equiv 1-\frac{\mathcal{F}}{100}$ for fidelity $\mathcal{F}$. Depending on the method and target fidelity, readout times in these systems may vary between tens and hundreds of $\us$.
Single qubit gates have been performed with errors $\varepsilon = 1.0(3)\cdot 10^{-6}$ for an rf $\pi/2$-pulse of $12\us$ pulse duration on a single ion \cite{Harty2014}.
Fast single qubit gates can be implemented with a pulsed laser trading off fidelity against speed, achieving $\varepsilon = 7\cdot 10^{-3}$ for $t_{\pi/2}=40\ps$ \cite{Campbell2010}.
Two qubit gate errors are $\varepsilon = 8(4)\cdot 10^{-4}$ at a gate time of $t_g=30\us$ \cite{Gaebler2016,Ballance2016,Baldwin2020} for laser gates, and $\varepsilon$ in the interval $[ 1.7\cdot 10^{-3},0]$ at a gate time of $t_g=740\us$ for microwave gates \cite{Srinivas2021}.
The fastest two-qubit gates achieved $\varepsilon = 2.2(3)\cdot 10^{-3}$ in $t_{g}=1.6\us$ \cite{Schafer2018}.

\subsection{Use case}

Various algorithms have been implemented on trapped ion systems, including Shor's algorithm and Grover's search algorithm \cite{Monz2015b,Figgatt2017}, demonstrations of error correction \cite{Negnevitsky2018,Egan2020}, analogue quantum simulations, such as the simulation of many-body dynamical phase transitions \cite{Zhang2017a} exceeding the capabilities of classical computers, as well as several VQE demonstrations \cite{Hempel2018,Nam2020,Foss-Feig2020}, for example estimating the ground state energy of $\mathrm{H_2}$, LiH and $\mathrm{H_2O}$.

\begin{figure}
\centering
\includegraphics[width=8cm]{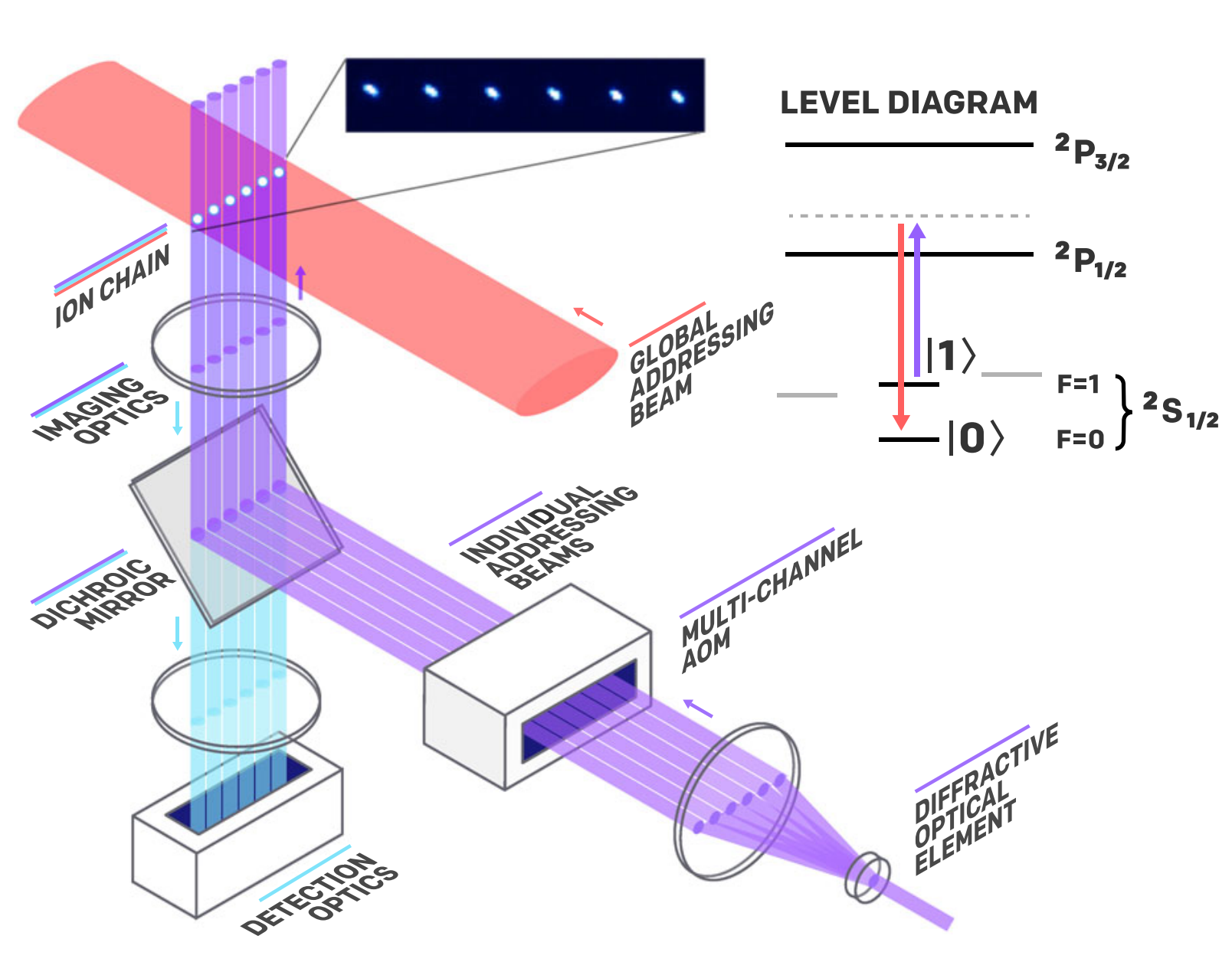}
\caption{\textbf{IonQ quantum computer based on a chain of trapped ions:} 
A high-numerical aperture lens allows both individual addressing and readout of the ions. 
A multi-channel AOM is used to modulate the amplitude, frequency and phase of the individual laser beams.
Inset: The qubits are encoded in the hyperfine ground-states $\up={^2\mathrm{S}_{1/2}},\mathrm{F}=1$ and $\dn={^2\mathrm{S}_{1/2}},\mathrm{F}=0$ of the trapped \yb ions. 
Gate operations are performed via a two-photon Raman process, coupling to the excited $^2\mathrm{P}$ states (purple and orange beams). 
Figure adapted from \cite{Nam2020}.}
\label{fig:ionq_qc}
\end{figure}

Fig.\ \ref{fig:ionq_qc} shows the ion trap quantum computer of IonQ, which is commercially accessible and was used to perform VQE on three individually addressable qubits encoded in a string of \yb ions to estimate the ground state energy of the water molecule \cite{Nam2020}.
The quantum circuit implementation for the energy-evaluation was optimised to take advantage of the asymmetric state measurement fidelities of the \up and \dn states, and the higher fidelity ($\varepsilon_{\phi=\pi/100}\lesssim 4\cdot 10^{-3}$) of partially entangling gates $\mathrm{XX(\phi)}$ ($\phi<\pi/2$) compared to full entangling gates  $\mathrm{XX(\pi/2)}$ ($\varepsilon\lesssim 4\cdot 10^{-2}$). The longest implemented circuit comprised 6 CNOT two-qubit operations.
An energy uncertainty close to the chemical uncertainty of $1.6\,\mathrm{mHa}$ was achieved (albeit in a minimal basis set), without a need for implementing error mitigation techniques such as Richardson extrapolation \cite{Richardson1927}.

\section{\label{sec:silicon}Silicon quantum computer}
Today’s digital age is enabled by the relentless progress and optimisation of semiconductor materials and technology. From an industrial standpoint, the use of well-established nanofabrication techniques for the development of quantum machines would be economically attractive to achieve large-scale systems. As discussed, some of these manufacturing techniques are already applied to superconducting and ion trap quantum platforms, and are expected to become central for the development of silicon-based systems, offering the prospect of integrating millions of qubits on chips at affordable manufacturing costs, akin to classical commercial electronics. Besides this technological motivation, silicon is a particularly suitable material for spin-based quantum devices from a performance viewpoint. Through isotopic purification, the only isotope bearing a nuclear spin ($^{29}$Si) in natural silicon can be nearly completely removed, making the silicon crystal a quasi-spin-noise-free environment for the qubit. This results in silicon spin-qubits having the longest coherence time among solid-state implementations.\\\indent 
Besides silicon, there exists a large variety of semiconductor systems currently under investigation for quantum computing applications~\cite{Atature2018,chatterjee2020semiconductor,Lieven2019,Bluhm}. The main differences lie in the type of material (e.g. natural or purified silicon, synthetic diamond, silicon carbide, heterostructures such as GaAs/AlGaAs, Si/SiGe or Ge/SiGe), the operational conditions (ranging from room temperature down to millikelvin temperature), the way each qubit is spatially confined within the material (e.g. gate-defined quantum dots, etched nanowires, atomic-size crystallographic defects, implanted dopant impurities), the way the qubit state is readout (e.g. electrical readout via charge sensors, or optical readout through photoluminescence), and the way the qubit state is manipulated (e.g. electron spin resonance via magnetic field pulsing, electric dipole spin resonance via electric field pulsing). Despite such diversity, a common denominator in most platforms is the choice of electron/hole spins as the two-level system embodying the qubits. The paradigmatic encoding is represented by a single spin in a static magnetic field with its two Zeeman-split energy levels representing the states $\ket{0}$ and  $\ket{1}$. Other implementations that have been explored include two-electron singlet-triplet qubits, three-electron charge-spin hybrid qubits and three-electron exchange-only qubits. Such rich ecosystem gives rise to significant performance variations among qubit implementations. The trade-offs can be many, including the robustness to specific noise sources and the ease of operation. The coherence times can range from few tens of nanoseconds in GaAs/AlGaAs quantum dots to few seconds in silicon dopants, and the single-qubit gate time can vary between sub-nanosecond and hundreds of nanoseconds in Si/SiGe quantum dots and silicon dopants, respectively.\\\indent 
In this Section, we are going to focus our attention on a particular type of semiconductor qubit system, which has been deployed for the realisation of the first spin-based quantum computer in the Cloud:  SPIN-2QPU~\cite{Last2020}, developed at QuTech (a collaboration between TUDelft and TNO). It consists of two single electron spin qubits in a double quantum dot (DQD) that is electrostatically defined by metallic gate electrodes deposited on top of an isotopically purified Si/SiGe heterostructure, as illustrated in Fig.~\ref{fig:si_device}~(a) and Fig.~\ref{fig:si_device}~(b). \\\indent
Similar to the other quantum processors discussed previously, spin-based machines must meet certain functional criteria. These include reliable initialisation to a known state, high fidelity projective readout of the final state, and qubit manipulation through high-quality single- and two-qubit gates. Let us see how SPIN-2QPU satisfies these criteria.
\subsection{\label{sec:init}Qubit initialisation \& readout}
\begin{figure}[]
\includegraphics[scale=0.45]{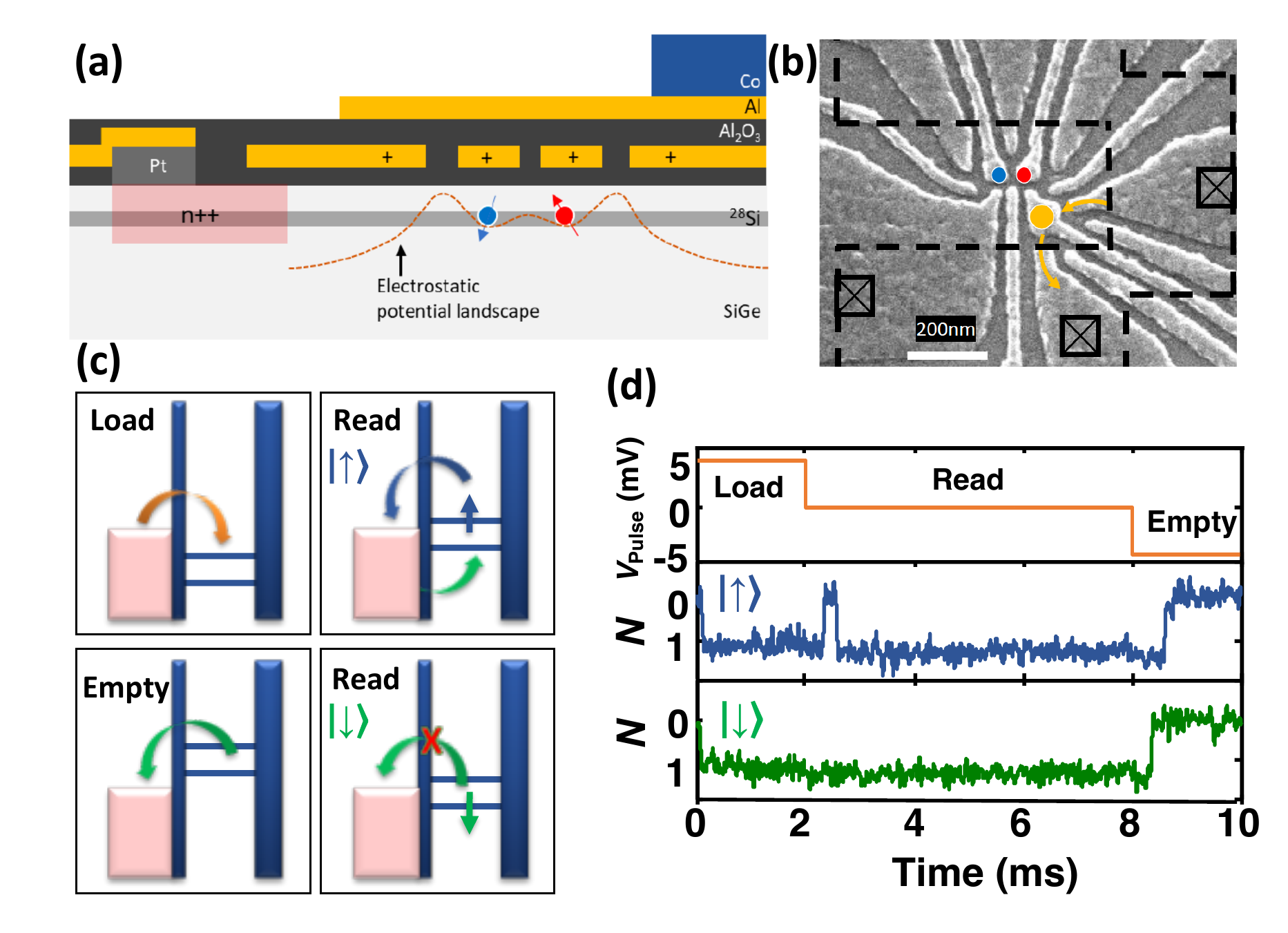}
\caption{(a) Schematic cross-sectional view of a DQD device used to control two spin qubits. Top metal gates (yellow areas) are used to tune the conduction band profile (dashed line) in the Si layer and isolate two electron spins. An n-type doped region of semiconductor (pink shaded area) is used as an electron reservoir tunnel-coupled to the left QD. A layer of cobalt (blue box) is deposited on top of the gate layers to generate a controlled magnetic field gradient across the DQD. (b) SEM micrograph of the DQD in (a). The aluminum gates are patterned with electron-beam lithography. The two qubits are formed under the gates highlighted with red and blue circles. The SET detector is formed under the gates highlighted in yellow. Gates that accumulate the electron reservoirs for the DQD and the SET are connected to Ohmic contacts and highlighted by crossed squares. Dashed lines indicate the region where the micromagnet is deposited. (c) Schematic diagram of energy levels for the left QD and the electron reservoir during the readout pulse sequence. Energy levels in the QD are Zeeman-split according to spin polarization.  (d) Pulsing sequence (top) for the spin readout and normalised SET signal for spin-up (middle) and spin-down (bottom) qubit states. Panels (a) and (b) are adapted from Ref.~\cite{Last2020}. Panels (c) and (d) are adapted from Ref.~\cite{Yang2013}.}
\label{fig:si_device}
\end{figure}

The readout of the qubit state is ultimately a measurement of the electron spin orientation. However, the magnetic moment of a single spin is exceedingly small and its direct detection quite difficult. By contrast, the detection of small displacements of single charges is routinely carried out in semiconductor devices. To this end, SPIN-2QPU uses a single-electron transistor (SET) capacitively coupled to the DQD, as shown in Fig.~~\ref{fig:si_device}~(b). Whenever a single electron leaves/enters the DQD, the SET produces a discrete jump in the value of its electric current caused by a change in its operation point.\\\indent 
Reading out the spin state is, therefore, a matter of making a so-called spin-to-charge conversion, whereby the electron is allowed to tunnel in or out the DQD in a way that depends on its spin state, equivalent to whether the qubit is in state $\ket{0}$ or $\ket{1}$ . As shown in Fig.~\ref{fig:si_device}~(c), the selection rule is energy-based. A single spin in one of the dots is capacitively coupled to the SET and tunnel coupled to a reservoir. After spin manipulation, the dot's energy level is tuned with a gate voltage pulse such that the Fermi reservoir lies between the two Zeeman-split spin states. If the electron is in state $\ket{\downarrow}$, it does not have enough energy to leave the dot, and there is no SET current change due to a lack of charge rearrangement. For a state $\ket{\uparrow}$, the electron can tunnel out of the quantum dot and into the reservoir, leading to a change in SET current until a new electron tunnels in and re-initializes the qubit to its ground state. Current traces for these two alternative scenarios are shown in Fig.~~\ref{fig:si_device}~(d). Note that in this system initialization can be seen as a by-product of readout, given that an electron with a known spin, i.e. $\ket{\downarrow}$, always resides in the dot at the end of the sequence.

\subsection{\label{sec:manip}Qubit manipulation}
Analogously to other qubit realisations, a spin-qubit requires independent rotations about the axes of the Bloch sphere (single-qubit gate), as well as rotations that are dependent on the state of another qubit (two-qubit gate), in order to form a set of universal quantum gates. Through a two-qubit gate, entangled states can be created when one of the two qubits starts in a superposition of states.\\\indent
SPIN-2QPU carries out single-qubit gate operations through electric dipole spin resonance (EDSR). It consists of a microwave modulated electric pulse delivered through a gate electrode that oscillates the electron wavefunction. This has the effect of rotating the electron spin whenever the electron experiences a time-varying magnetic field resonant with its Zeeman splitting. This requires the presence of a synthetic spin-orbit field obtained through a local magnetic field gradient in the DQD, which is engineered by depositing a cobalt micromagnet on top of the device gate layer (see Fig.~~\ref{fig:si_device}~(a)). The amplitude of the EDSR pulse controls the spin vector’s rotation frequency around the Bloch sphere, its phase controls the rotation axis, and its duration controls the rotation angle. The frequency of the pulse allows one to select which qubit is manipulated, given that each electron experiences a slightly different magnetic field due to the different position within the DQD.\\\indent
SPIN-2QPU carries out two-qubit gate control via modulation of the exchange interaction. The idea is to quickly turn on the tunnel coupling between two neighbouring spins by applying a gate-voltage pulse that lowers the tunnel barrier between their corresponding quantum dots, so that the electron wavefunctions overlap. Such overlap leads to an exchange interaction between the spins, which can be exploited for conditional gate operations.

\subsection{\label{sec:oper}Operational conditions \& performance indicators}
The readout protocol is effective if the qubit energy levels are separated by at least a few times the thermal energy. This is ultimately the reason why SPIN-2QPU and similar semiconductor-based quantum systems need to be operated at dilution refrigerator temperature ($T$) and in the presence of an external static magnetic field ($B$). Typical conditions require $B\approx$~1~T and $T\approx$~50~mK. The duration of the readout sequence is ultimately determined by the tunnelling rate between the DQD and the reservoir, as well as by the bandwidth of the SET detector. SPIN-2QPU’s readout duration is $\approx$~300~$\mu$s per qubit and its readout fidelity is approximately 85\%.\\\indent
Given a single-qubit gate duration of approximately $250$~ns and a phase coherence time of at least 6~$\mu$s, SPIN-2QPU achieves single-qubit fidelities in excess of 99.0\%. As for 2-qubit operations, the only allowed native gate is CZ. Hence, other gates like CNOT and SWAP have to be decomposed into CZ operations in combination with single-qubit rotations. This comes at the expense of fidelity and operational time. A detailed benchmark for CZ is ongoing. Preliminary data show gate duration of around 150 ns and fidelity in excess of 90\%, but this latter figure is likely to be a conservative underestimate at this stage.

\subsection{\label{sec:use}Use case}
At present, we are not aware of VQE simulations carried out with SPIN-2QPU or any other semiconductor qubit system, possibly due to the limited qubit count. By contrast, 2D arrays of semiconductor QDs have been used for analog simulations of magnetic and insulating materials by spatially engineering Hamiltonians onto the array~\cite{Hensgens2017,Dehollain2020}. It is, however, useful to report that it has been possible to run digital algorithms of different kinds (Deutsch–Josza and Grover) on the SiGe processor that QuTech used to prototype SPIN2-QPU~\cite{Watson2018}. This ultimately casts a positive light for future uses of semiconductor machines in computational chemistry. 

\begin{table*}[t]
\begin{center}
\begin{tabular}{ |p{2.3cm}|p{2.3cm}|p{2.3cm}|p{2.3cm}|p{2.3cm}|p{2.3cm}|p{2.3cm}|  }
 \hline
 \textbf{Manufacturer} & \textbf{Platform} & \textbf{Cloud access} &\textbf{Max \# qubits} &\textbf{ Gate fidelity (1-qubit, 2-qubit) }&\textbf{Max QV} &\textbf{Simulated molecules}\\
 \hline
 IBM   & Superconducting    &IBM Quantum Experience (Open access) & 15 (Melbourne)&  99.97\%, 99.16\% (Santiago) & 32 (Santiago) &H$_2$, LiH, BeH$_2$, NaH, KH, RbH\\\hline 
 IonQ& Trapped Ions  & Microsoft Azure or Amazon Bracket & 11 &  99.50\%, 97.50\% & not published & H$_2$O \\\hline
 QuTech   & Silicon    & Quantum Inspire & 2 (Spin2-QPU) &  $\approx$~99\%, $\approx$~90\% &not published &none\\\hline 
 Google&   Superconducting  & Google Quantum AI &53 (Sycamore)& 99.85\%,
99.35\% & not published & H$_2$N$_2$, H$_6$, H$_8$, H$_{10}$, H$_{12}$ \\\hline
 Rigetti   & Superconducting    & Rigetti Quantum Cloud  & 31 (Aspen-8) &  99.8\%, 95.9\% & 8 (Aspen-4) & NaH, H$_2$\\\hline 
 Honeywell& Trapped Ions  & Microsoft Azure or Amazon Bracket &10 (H1)&99.97\%, 99.5\% &128 & none \\\hline
\end{tabular}
\caption{Quantum computing hardware in the cloud (a non-exhaustive selection). Wherever more than one QPU is available, the relevant machine is indicated within brackets. Simulated molecules column denotes experiments run with any quantum machine from the relevant manufacturer, not necessarily one of those listed. IBM hardware considered is limited to Open Access services. Google cloud services are limited to emulators at present, although the reported chemical simulations have been performed with proprietary physical hardware.}
  \label{tab:1}
\end{center}
\end{table*}
\section{\label{sec:disc}Discussion}
In recent months significant attention has been drawn to superconducting quantum hardware because a team at Google achieved a much anticipated milestone, namely quantum supremacy~\cite{Arute2019}. By quantum supremacy, it is meant that a quantum computer is able to produce the solution to a computational problem that would be otherwise impossible in a reasonable time with a classical machine. Google scientists achieved this with a 53-transmon-qubit processor (Table~\ref{tab:1}) by showing efficient sampling of random quantum circuits. While this result is of primary importance for the field as a whole, the problem tackled did not bear any relevance to molecular chemistry. Therefore, with regard to this type of problem, a quantum advantage is yet to be demonstrated. However, in a more recent study~\cite{Arute2020_1}, the Google team used the same quantum processor for chemical simulations, as discussed in Section~\ref{sec:super:use}. They demonstrated the most complex ground state simulation to date with as many as 24 spin-orbitals mapped onto 12 qubits. Although these calculations are relatively straightforward with a conventional supercomputer, they represent a significant advance of the state-of-the-art in quantum computing power, as the number of qubits used and orbitals simulated in prior experiments was no more than six~\cite{Kandala2017}. While Google's quantum hardware is scheduled to be deployed onto cloud services imminently, there is already a variety of tools made available by Google scientists to experiment with emulated hardware tailored for applications in molecular chemistry\cite{mcclean2019openfermion}. As for superconducting hardware readily available in the cloud, one has to currently turn to IBM or Rigetti, see Table~\ref{tab:1}. IBM has about a dozen QPUs in the cloud, arguably the most extensive offer yet. Just through its Open Access service, the community can access eight machines with qubit counts ranging from 1 to 15 and QV ranging from 8 to 32. The most powerful QPUs with qubit counts up to 65 and QV up to 32 are available for business clients via Premium Access. A recent breakthrough has led to QV=64 for a new 27-qubit system not yet available in the cloud. IBM scientists were among the pioneers in exploiting QPUs for molecular chemistry applications (see Table~\ref{tab:1})~\cite{Kandala2017}. More recently, they have also shown that improved simulation accuracy can be obtained by adopting error mitigation techniques at algorithmic level~\cite{Kandala2019}. This is important because it can be used to enhance the computational power of a processor without any hardware modification.\\\indent
Quantum machines based on trapped ions have progressed very quickly in the past year alone. While devices used for digital quantum computing typically have a lower qubit count than their superconducting counterparts, analogue quantum simulation has been performed on strings containing up to 53 qubits \cite{Zhang2017a} and single qubit operations have been performed in devices containing up to 79 qubits \cite{Wright2019}.
Due to superior gate fidelity and qubit-to-qubit connectivity, the quantum volume of ion trap processors is outperforming superconducting devices even for smaller numbers of qubits.
Recently, corporate research teams at IonQ and Honeywell have made QPUs available through the wider cloud services of Amazon and Microsoft, see Table~\ref{tab:1}. Honeywell's QPU shows the largest volume to date, i.e. QV=128. Both Honeywell and IonQ have recently announced  the imminent launch of  upgraded QPUs with significantly enhanced QV values.
Trapped ion machines have also been used for molecular chemistry simulations\cite{Hempel2018,Nam2020,Shen2017}. The most complex molecular simulation performed to date with trapped ions is the evaluation of the binding energy of the water molecule with a 3-qubit QPU from IonQ~\cite{Nam2020}, as discussed in Section~\ref{sec:ion}.\\\indent
The 2-qubit silicon quantum processor made by QuTech is the only spin-based system in the cloud. The service through which it is accessible, the platform~\href{http://www.quantum-inspire.com}{Quantum Inspire}, also provides a more powerful alternative based on a 5-qubit superconducting QPU. Silicon SPIN2-QPU has been the latest to be deployed (April 2020) and is not yet fully characterised, hence only approximate fidelities are quoted in Table~\ref{tab:1}. Although no chemical simulations have been attempted yet, one should expect that the semiconductor community will soon fill this gap. The modest qubit count should not be an insurmountable impediment if one considers that early 2-qubit QPUs were successfully used to simulate diatomic molecules~\cite{OMalley2016,Colless,Shen}. Undoubtedly, Si-based machines have yet to cover much ground before becoming realistic competitors of the other two major platforms. For example, high-fidelity single- and two-qubit gates have only recently been achieved and are not yet on par with those of the other hardware platforms~\cite{chatterjee2020semiconductor}. Furthermore, qubit variability due to atomic level defects in the material and its interfaces is an issue that currently hampers scalability. Nonetheless, the interest around these devices is justified by the fact that in principle they can be manufactured with industrial CMOS technology, and have the smallest qubit footprint~\cite{gonzalezzalba2020scaling}. This bodes well for future upgrades of such systems towards the million-qubit-machines needed for useful applications. Finally, note that there exists another type of silicon QPU based on photonic technology (as opposed to spins) with two systems accessible via cloud services~\cite{Xan, Bristol}.

\section{\label{sec:conc}Conclusion and Outlook}
A lot of theoretical and experimental ground has been covered since the early 80s, when Feynman proposed to use controllable quantum devices for computational problems in chemistry and physics. There are now dozens of small-scale quantum computers in the cloud and many more in academic and corporate laboratories worldwide. The electronic structures of simple molecules ranging from diatomic systems to chains of a dozen atoms has been determined with several QPU incarnations.\\\indent
In this Article, we have discussed the hardware of the most popular types of quantum computers, for which we have summarised the main techniques for physical encoding, manipulation and readout of quantum information. We have paid particular attention to the machines that the reader could easily access via cloud services, i.e. superconducting-, trapped ion- and silicon-based processors. For these, we have described the main performance specifications and operational conditions. Our target has been to highlight to what extent these early prototypes have been employed for chemistry simulations. The underlying message is that, despite relentless progress, none of the machines built thus far is yet advantageous to a chemist, if compared to classical computational methods. What needs to happen to change this?\\\indent
In order to achieve a sizable quantum advantage in computational chemistry with NISQ machines, the coordinated efforts between quantum hardware and quantum algorithm developers will need to continue if not intensify. Hardware improvements in terms of qubit count, qubit connectivity, quantum gate speed and fidelity, as well as overall QPU volume will be a central focus for years to come. These advances will be essential to bring quantum simulation run-times down to practical length-scales~\cite{elfving2020quantum}. However, recent breakthroughs~\cite{Kandala2019, Arute2020_1} have also shown that tailoring algorithms to the specific quantum hardware available in combination with error mitigation techniques could be important for accurate chemical computation on near-term machines. Particularly, restrictions to realisable gates inherent to NISQ processors could be bypassed with ad-hoc compilation methods.\\\indent
Beyond the NISQ era, i.e. without today's limitations due to noise, there will be the possibility of taking full advantage of the computational speed-up of quantum systems. QEC protocols will have to be reliably implemented to produce such step change. During this transition, a risk to be avoided will be that today's capability restrictions, rather than being lifted altogether, will be merely transferred from the quantum layer onto the classical control layer~\cite{martinis2020information}. There are two complementary considerations to this potential problem. Firstly, QEC will require fast feedback between measurement and control, and communication latency may become an issue. If there is a sizable physical distance between the quantum hardware and the classical control hardware, which is likely for cryogenic QPUs, delays in the communication lines may pose a synchronization challenge if they become of the same order as the gate time. Secondly, a computational bottleneck may occur in handling error correction cycles for large number of physical qubits. For example, a QPU with a million qubits corrected with cycles of 1~$\mu$s will require classical information processing at a bandwidth of 1 Tbit/s. If both latency and bandwidth issues are to be solved at once, novel ultra-low-power cryogenic RAM and CPU may need to be developed, so that they could sit near or within the same chip of a cryogenic QPU without generating detrimental heat loads~\cite{gonzalezzalba2020scaling}.
\\\indent
We believe that the challenges described do not represent a fundamental roadblock towards large-scale fault-tolerant quantum computing. However, they do pose significant engineering hurdles that will require synergies between quantum and electronic engineers, as well as quantum software developers and end users. We hope that this Article will trigger the curiosity of theoretical and quantum chemists in trying out the available cloud machines, get involved into the ongoing conversation and, eventually, steer quantum systems development to the benefit of their scientific agenda.\\

\begin{acknowledgments}
We wish to thank N. Samkharadze for useful discussions. AR acknowledges the support of the UK Government Department for Business, Energy and Industrial Strategy through the UK National Quantum Technologies Programme, as well as support from a UKRI Future Leaders Fellowship (MR/T041110/1). VMS acknowledges funding from Christ Church College, Oxford. PGB and MW acknowledge support from the EPSRC Hub in Quantum Computing and Simulation (EP/T001062/1).
\end{acknowledgments}

\bibliography{main_ref}

\end{document}